# Effects of nanoparticles on murine macrophages

M Chevallet[1], C Aude-Garcia[1], C Lelong[1], S Candéias[1], S Luche[1], V Collin-Faure[1], S Triboulet[1], D Diallo[1], H Diemer[2], A van Dorsselaer[2] and T Rabilloud[1]

[1]CEA/DSV/IRTSV, laboratoire de Biochimie et Biophysique des Systèmes Intégrés, Unité Mixte CNRS UMR5092, Université Joseph Fourier - Grenoble, CEA Grenoble, 17 rue des martyrs, 38054 Grenoble, France.

[2]IPHC, Institut Pluridisciplinaire Hubert Curien, CNRS UMR7178, Université Louis Pasteur – Strasbourg I, France.

E-mail: thierry.rabilloud@cea.fr

**Abstract.** Metallic nanoparticles are more and more widely used in an increasing number of applications. Consequently, they are more and more present in the environment, and the risk that they may represent for human health must be evaluated. This requires to increase our knowledge of the cellular responses to nanoparticles. In this context, macrophages appear as an attractive system. They play a major role in eliminating foreign matter, e.g. pathogens or infectious agents, by phagocytosis and inflammatory responses, and are thus highly likely to react to nanoparticles. We have decided to study their responses to nanoparticles by a combination of classical and wide-scope approaches such as proteomics. The long term goal of this study is the better understanding of the responses of macrophages to nanoparticles, and thus to help to assess their possible impact on human health.
We chose as a model system bone marrow-derived macrophages and studied the effect of commonly used nanoparticles such as $TiO_2$ and Cu. Classical responses of macrophage were characterized and proteomic approaches based on 2D gels of whole cell extracts were used. Preliminary proteomic data resulting from whole cell extracts showed different effects for $TiO_2$-NPs and Cu-NPs. Modifications of the expression of several proteins involved in different pathways such as, for example, signal transduction, endosome-lysosome pathway, Krebs cycle, oxidative stress response have been underscored. These first results validate our proteomics approach and open a new wide field of investigation for NPs impact on macrophages.

## 1. Introduction

Nanoparticles (NPs) are increasingly used for developing new tools for biotechnology, as well as manufacturing widely-used products. They offer interesting properties, regarding for example electromagnetic characteristics, mechanical properties, chemical and/or biological reactivity. Nanoparticles can be used for medical imaging and drug delivery, as well as for electronic devices [1-5]. Moreover, many household products contain NPs of metal oxides, such as sunscreens, cosmetic products, or clothes [6, 7]. New tools have also been developed for biotechnology and life sciences, using metallic NPs as well as nanotubes, nanowires or quantum dots [8]. This massive development of nanotechnologies requires studies to assess potential associated biological risks.

Thus, there is an important corpus of literature assessing the effects of nanoparticles on living organisms. This literature can be split in two types of studies: on the one hand, there are in vivo studies on animal models where macroscopic results can be obtained [9-12]. However, because of the short lifespan of laboratory animals, long-term effects are difficult to derive. Moreover, the in vivo systems are so complex that it is almost impossible to get mechanistic data that would help to predict long term effects. On the other hand, there are in vitro studies on cells [6, 8, 13-17]. This simplified system can be used to derive mechanistic data. However, most of this literature focuses on a few cellular parameters (survival, oxidative stress, cytokine productions) so that very little is known on the possible responses elicited by nanoparticles on cells.

In order to bring new, mechanistic knowledge, we have decided to use a wide-scope, proteomics approach. As a cell type, we have chosen the macrophage, i.e. the specialized cell type that will scavenge bacteria or viruses [18] as well as any type of particulate material, nanoparticles [19]. Indeed, macrophages have been shown to be key players in the toxicity of asbestos fibers and silica nanoparticles, and thus in the etiology of asbestosis and silicosis, by the virtue of their stress response to the phagocytosis of these particles [20].

Thus, we have decided to apply proteomics to investigate the response of macrophages to TiO2 considering that this NP is widely produced and therefore, potential widespread exposure may occur during both manufacturing and use. We also compared the effects of $TiO_2$ with Cu, which at the difference of $TiO_2$ is a redox metal and can release cupric ions which can be toxic for cells. Moreover, Cu-NPs are also widely-used NPs because of their antibacterial properties. This work has shown changes in the central cellular metabolism, as well as a peculiar oxidative stress response.

## 2. Materials and methods

*2.1. Mice*
C57BL/6 mice were bred in the animal facility of the Commissariat à l'Energie Atomique, Grenoble, France. For hind legs recovery, 10-20 week-old mice were sacrificed by cervical dislocation in accordance with Atomic Energy Commission Care and Use Committee.

*2.2. Primary macrophages*
Unless otherwise stated, all chemical reagents and cell culture products were purchased from Sigma-Aldrich (Saint Quentin Fallavier, France). Murine primary macrophages were generated from bone marrow as described [21]. Briefly, the bone marrow cavities of femurs and tibiae were flushed several times with BM20 medium (DMEM supplemented with 10% FCS, 20% L929 cell-conditioned medium as a source of M-CSF, 5% horse serum, 1% glutamine and 1% sodium pyruvate). 40-60 x $10^6$ cells were recovered per mouse and cultured in 50 ml of the same medium. At day 7, the supernatant was removed. Adherent cells were incubated for 10 min at 4°C in cold 0.02%EDTA in PBS. Cells were then recovered and transferred in 50 ml BM10 medium (same as BM20 but with 10% L929 cell-conditioned medium) for 3 additional days. 20-30 x $10^6$ cells were generated at the end of the differentiation process and most of the resulting cells (> 90%) were CD11b and CD14 positive, as assessed by flow cytometry. These mature macrophages, derived from bone marrow exactly as occurs in vivo, are therefore a good model of all the macrophages that exist in vivo, alveolar as well as peritoneal, dermal or hepatic.

*2.3. Flow cytometry analysis*
For the analysis of cell surface protein expression, bone marrow-derived macrophages were stained for 30 min at 4°C with conjugated antibodies in PBS containing 3% fetal calf serum and 0.16% sodium azide. Anti-CD11b and PE-conjugated anti-CD14 were from CliniScience (Montrouge, France), FITC-conjugated anti-CD86 (B7-2) and PE-conjugated anti $IA^b$ were from BD Pharmingen (le Pont de Claix, France). Data acquisition and analysis were performed with a FacsCalibur flow cytometer equipped with CellQuest software (Becton Dickinson).

*2.4. Preparation of nanoparticle suspensions*
Titanium oxide ($TiO_2$, anatase, < 25nm, 99.7% purity) and Copper (Cu, < 50nm, 99.5% purity) were purchased from Sigma-Aldrich (France). For their sterilization, the NPs were suspended in 95% ethanol, at a concentration of 10 mg/ml. After 1h incubation at room temperature, the NPs were diluted to a final concentration of 1 mg/ml in polyvinylpyrrolidone 40 (PVP40) 0.2%, sonicated in a bath sonicator for 1h and stored at 4°C. PVP40 increases the dispersions of the NPs but there is still some aggregation in the culture media. Characterization of the particle size distribution was performed by Dynamic Light Scattering measurements (DLS) which indicated a Z-average of 318.9 nm for TiO2 NPs and of 427.8 nm for Cu NPs after 24h incubation in culture media. We chose to perform this polymer coating to mimic the opposite phenomena occurring when nanoparticles are in contact with a biological fluid, whether it is pulmonary surfactant, saliva, lymph or blood. On the one hand the salt concentration present in all these fluids promote aggregation, so that nanoparticles are never as dispersed in biological media as they can be in air or in pure water. On the other hand, the polymeric materials present in biological fluids (polysaccharides, proteins) will act as coating agents and limit particle aggregation. We have used PVP 40 for its lack of interference with our analyses, while being perfectly biocompatible and able to limit particles aggregation.

*2.5. Exposure of macrophages to nanoparticles*
Cells ($5 \times 10^6$) were cultured in Petri dishes and treated for 24h with either 100 µg/ml $TiO_2$ or 10 µg/ml Cu-NPs. Two types of negative controls were done with and without PVP40.

*2.6. Two-dimensional electrophoresis*
Primary macrophage cells were harvested by scraping, rinsed 2 times in phosphate-buffered saline and pellets were suspended in homogeneization buffer (0.25 M sucrose, 10 mM Tris-HCl, pH 7.5, 1 mM EDTA). A buffer volume approximately equal to the packed cell volume was used. The suspension was transferred to polyallomer ultracentrifuge tubes, and cells were lysed by the addition of 4 volumes (respective to the suspension volume) of 8.75 M urea, 2.5 M thiourea, 25 mM spermine and 13 mM Tris carboxyethylphosphine. After 1 hour at room temperature, the extracts were centrifuged (30 min at 200,000 x *g*). The supernatant was collected and the protein content was determined by the Bradford protein assay using bovine serum albumin as a standard. The protein extracts were stored at −20 °C. The first dimension of electrophoresis was performed with immobilized pH gradients for isoelectric focusing. Linear 4–8 pH gradients were used. Home-made pH gradient plates were cast and cut into 4-mm-wide strips [22, 23]. A total of 150 µg of proteins were diluted in 560 µl of rehydratation buffer (7 M urea, 2 M thiourea, 4% CHAPS, 0.4% ampholytes, 100 mM dithiodiethanol). The samples were applied onto the strips by in-gel rehydration overnight using a thiourea-urea mixture as denaturing agent [24]. IEF was carried out for 60,000 Vh at a maximum of 3000 V using the Multiphor II system (Amersham-Pharmacia, Sweden). Strips were then equilibrated for 20 min in 0.15 M BisTris/0.1 M HCl, 6 M urea, 2.5% SDS, 30% glycerol. Strips were placed on top of a SDS-polyacrylamide gel. After migration, the gels were stained with silver [25]. Expression ratios were estimated by image analysis of pairs of gels with the Delta2D software (DECODON GmbH, Greifswald, Germany). Spots of interest (ratio >1.5) were cut and analysed by MS/MS.

*2.7. MS/MS analysis*
In gel digestion was performed with an automated protein digestion system, MassPrep Station (Waters Corp., Milford, USA). The gel plugs were washed twice with 50 µL of 25 mM ammonium hydrogen carbonate ($NH_4HCO_3$) and 50 µL of acetonitrile. The cysteine residues were reduced by 50 µL of 10 mM dithiothreitol at 57°C and alkylated by 50 µL of 55 mM iodoacetamide. After dehydration with acetonitrile, the proteins were cleaved in gel with 10 µL of 12.5 ng/µL of modified porcine trypsin (Promega, Madison, WI, USA) in 25 mM $NH_4HCO_3$. The digestion was performed overnight at room temperature. The generated peptides were extracted with 60% acetonitrile in 5% acid formic. NanoLC-MS/MS analysis was performed using an Agilent 1100 series nanoLC-Chip system (Agilent Technologies, Palo Alto, USA) coupled to an HCT Plus ion trap (Bruker Daltonics, Bremen, Germany). The chip was composed of a Zorbax 300SB-C18 (43 mm × 75 µm, with a 5 µm particle size) analytical column and a Zorbax 300SB-C18 (40 nL, 5 µm) enrichment column. Elution of the peptides was performed at a flow rate of 300 nL/min with a 8-40% linear gradient (solvent B, 98%ACN/0.1%FA) over 7 first minutes. The voltage applied to the capillary cap was optimized to -

1750V and the mass range was 250-2500 *m/z*. For tandem MS experiments, the system was operated Data-Dependent-Acquisition (DDA) mode with automatic switching between MS and MS/MS. The three most abundant precursor ions were selected to be further isolated and fragmented. The MS/MS scanning was performed in the ultrascan resolution mode at a scan rate of 26.000 *m/z* per second. A total of six scans were averaged to obtain an MS/MS spectrum. The system was fully controlled by the ChemStation (v 5.3) and EsquireControl (Rev B.01.03) software (Agilent Technologies and Bruker Daltonics, respectively). For protein identification, the MS/MS data were interpreted using the MASCOT 2.2.0. algorithm (Matrix Science, London, UK). Spectra were searched with a mass tolerance of 0.2 Da in MS and MS/MS modes, allowing a maximum of one trypsin missed cleavage. Carbamidomethylation of cysteine residues and oxidation of methionine residues were specified as variable modifications. Following the guidelines for proteomic data publication [26, 27], and to avoid consideration of poor quality data, filtering criteria based on probability-based scoring of the identified peptides have been taken into account for high confidence identification and a decoy database was generated and searched to determine false positive rate in protein identification [28]. The Scaffold software (v.3, Proteome software Inc., Portland, OR) was used in order to validate identification using the following criteria: spectra were searched against a target-decoy version of the UniProtKB/SwissProt database restricted to Mammalia downloaded in January 2010 (1028424 target and decoy entries). Protein identifications were validated when at least two peptides with Mascot identity scores of greater than 10 were detected.

## 3. Results and discussion

Before adding the NPs onto the bone marrow-derived macrophages, we checked by flow cytometry that these cells were mature macrophages, expressing both CD11b and CD14. Moreover, the cells obtained could be activated by LPS, as shown by release of TNFα and NO (data not shown). Using fluorescent beads, we also verified by flow cytometry that they were phagocytic. The concentration of NPs used was determined beforehand by viability tests (LD20). These doses are not intended to represent doses for a given body weight and a typical exposure. However, they are legitimate for a mechanistic study to exemplify what deleterious mechanisms can be at play during pathogenesis. Moreover, it should be kept in mind that doses per body weight are also an approximation, and that "overloads" can exist locally in vivo, e.g. in the dermis just below a wound in the case of cutaneous exposure, and that a single dose in vitro can also mimic the accumulation of low doses occurring vivo during the lifespan of a macrophage, as macrophages are known to accumulate nanoparticles [10]. Furthermore, in the case of cancers where a single event can lead to the pathology, this "hotspot" model is relevant.

We performed two comparative proteomics analysis: control versus $TiO_2$ treated cells and control versus Cu treated cells.

*3.1. Analysis of total proteome changes after NPs-TiO$_2$ exposure*

A total of 9 protein spots with an increased intensity as well as 1 spot with a decreased intensity after $TiO_2$ treatment could be detected (Figure 1, Table 1). They were excised from gels and analyzed by MS. All these spots were identified. $TiO_2$ treatment altered metabolism as showed by increase of spots corresponding to malate deshydrogenase, aldehyde deshydrogenase and glucosidase. We also observed an increase in VDAC (mitochondrial porin), peroxiredoxin 4 (Prx 4), normal form and proteasome alpha6 subunit, which are usually seen in various stress conditions. All these proteins belong to the "déjà vu in proteomics" [29, 30].

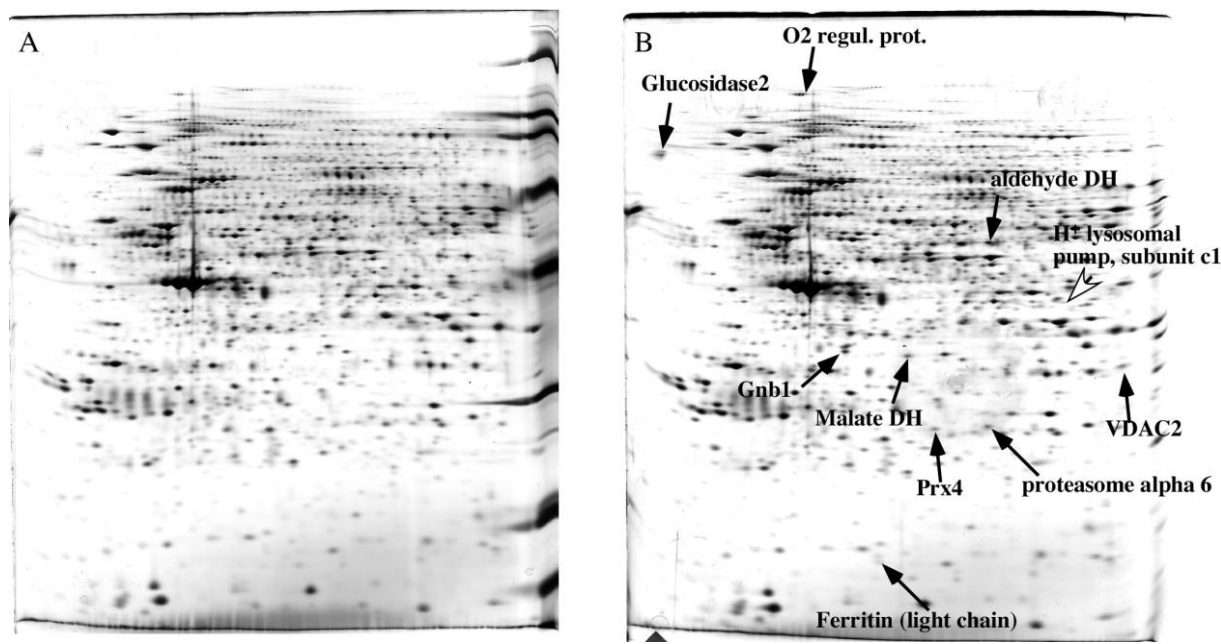

**Figure 1.** Comparison of the total extract patterns of control primary macrophages and $TiO_2$-NPs treated cells. Primary macrophages were treated with 100 μg/ml $TiO_2$-NPs for 24h. 150 μg of total proteins were separated by 2-D and detected with silver staining as described in Section 2. Gel images were analyzed by Delta-2D. (A) control cells, (B) $TiO_2$-NPs treated cells. Solid and empty arrows point at increased (≥1.50-fold) and decreased (≤0.67) proteins, respectively.

Cells responded to such a "light stress" by increasing general metabolism and some classical proteins. We identified some more original proteins: Gnb1, a G-protein implicated in signal transduction, ferritin (light chain) implicated in iron homeostasis and $O_2$-regulated protein, which is activated by hypoxia. This suggests that $TiO_2$ treatment may induce hypoxia-like conditions. At last, we noted that the subunit c1 of $H^+$ lysosomal pump was decreased. It is not surprising that the lysosomal pathway is altered by NPs phagocytosis as it is known that NPs can use the endosome-lysosome pathway [19].

*3.2. Analysis of total proteome changes after NPs-Cu exposure*

The same process was repeated for the 6 protein spots increased after Cu treatment and for the one with decreased intensity (Figure 2, Table 1). Again, Cu treatment increased some proteins spots belonging to metabolism and "déjà vu" in proteomics, such as ATPase subunit d, prohibitin and HSP70 (chaperone). Cells responded to stress by increasing metabolism, energy production and by increasing chaperones for protein refolding. More interestingly, an increase in heme oxygenase and peroxiredoxin1 (Prx 1), oxidized form was a signature of an oxidative stress response. In basal condition, this spot is very weak but increased almost 4 times upon Cu treatment. The oxidized spot level of Prx is a well-known indicator of the oxidative injury to the cell [31]. Heme oxygenase is generally induced when the level of free heme increases and this is associated to an oxidative stress. Moreover, heme oxygenase appears to be a potential defense against oxidative stress [32]. This result is consistent with a metal redox stress. In addition, two proteins implicated in endosome-lysosome pathway were also modified. MVB prot 4b was decreased and the d1 subunit of $H^+$ lysosomal pump was increased.

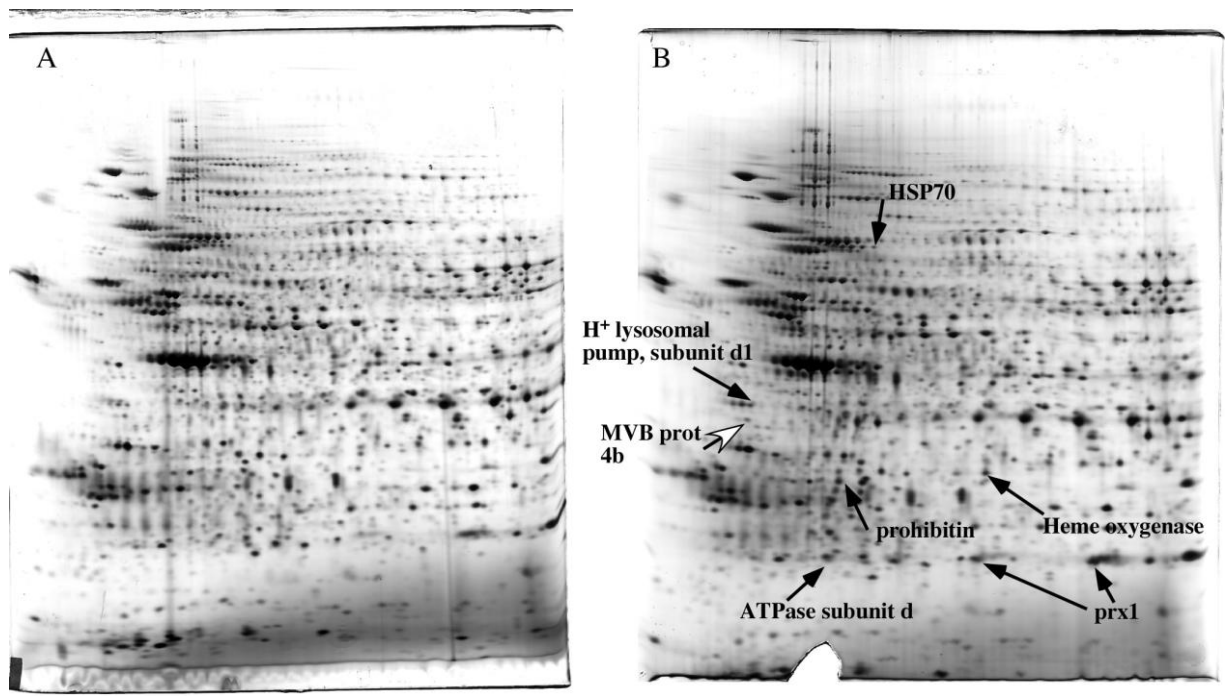

**Figure 2.** Comparison of the total extract patterns of control primary macrophages and Cu-NPs treated cells. Primary macrophages were treated with 10 μg/ml Cu-NPs for 24h. 150 μg of total proteins were separated by 2-DE and detected with silver staining as described in Section 2. Gel images were analyzed by Delta-2D. (A) control cells, (B) Cu-NPs treated cells. Solid and empty arrows point at increased (≥1.50-fold) and decreased (≤0.67) proteins, respectively.

*3.3. Comparison of total proteome changes after TiO$_2$-NPs and Cu-NPs exposure*
When we compared the two lists of proteins, we observed that there was no protein in common (table1). In both cases, we found metabolic proteins and classical stress proteins. We found more of these groups of proteins after TiO$_2$ treatment than after Cu treatment. However, Cu treatment appeared to be more toxic as it clearly induced oxidative stress. With TiO$_2$ treatment, we just noted an increase in the normal form of Prx 4, while Cu induced an important increase in oxidized forms of Prx 1 and of heme oxygenase. Peroxiredoxins are enzymes detoxifying reactive oxygen species, but they can get inactivated by oxidation in case of severe oxidative stress [31]. Presence of oxidized form is thus a signature of a deleterious oxidative stress. It is not surprising that Cu, which is a redox metal, generated such an oxidative stress. After TiO$_2$ treatment, ferritin was increased, possibly to protect the cell against free iron and oxidative stress. In addition, the two NPs affected the endosome-lysosome pathway even if we did not find the same proteins and found that two subunits of the same complex (H$^+$ lysosomal pump) varied in an opposite way.

The results obtained were different in terms of proteins but consistent in terms of function (metabolism, core stress response and endosome-lysosome pathway). Cu-NPs appeared in our conditions to be more toxic than TiO$_2$-NPs, and responsible of an oxidative stress. We observed with proteomics specific effects of Cu or TiO$_2$-NPs treatment. The fact that we did not observe the same list of proteins with these two different NPs showed that NPs toxicity is depending primarily of the nature of the NPs.

**Table 1.** Ratio NPs/control of the spots intensities measured by Delta 2D software on silver stained 2D gels.

| TiO$_2$ | | Cu | |
|---|---|---|---|
| VDAC 2 | 2.420 | HSP 70 | 2.800 |
| Aldehyde deshydrogenase | 2.240 | Prohibitin | 1.733 |
| Prx 4, normal form | 1.793 | ATPase subunit d | 1.700 |
| Proteasome α 6 | 1.763 | | |
| Glucosidase | 1.737 | | |
| Malate deshydrogenase | 1.498 | | |
| Ferritin (light chain) | 5.630 | Prx 1, oxidized form | 3.920 |
| O$_2$ regulated protein | 1.892 | Heme oxygenase | 3.250 |
| Gnb1 | 1.468 | H+ lysosomal pump d1 | 1.670 |
| H+ lysosomal pump c1 | 0.580 | MVB Prot 4b | 0.652 |

## 4. Perspectives
From the above, proteomics appears to be a relevant approach for mechanistic studies of cellular responses to nanoparticles. However, it must be kept in mind that such mechanistic studies are hypothetical and must be ultimately verified on relevant in vivo models. Nevertheless, this preliminary study showed specific effects of Cu-NPs and TiO$_2$-NPson murine bone-marrow macrophages. In this study, we looked at total cell extracts and thus only at more abundant and soluble proteins. To get further insights into the cell response to NPs, it will be of great interest to investigate different compartments and/or sub fractions of the cells. Such studies will bring more information on NPs toxicity pathways. We also consider to investigate toxicity of other NPs of different size and nature, to determine if a general pattern emerges or if each NP elicits a unique response.

## 5. Acknowledgments
This project belongs to the NanoBioMet consortium and is partially financed by the Toxicology transversal program.